\documentclass[aps,pre,twocolumn,floatfix,superscriptaddress,longbibliography]{revtex4-1}
\usepackage{graphicx,txfonts,bm}
\usepackage[colorlinks,citecolor=magenta,linkcolor=blue]{hyperref}

\begin{document}

\title{Combining Bayesian reconstruction entropy with maximum entropy method for analytic continuations of matrix-valued Green's functions}

\author{Songlin Yang}

\author{Liang Du}
\email{liangdu@gxnu.edu.cn}

\author{Li Huang}
\email{lihuang.dmft@gmail.com}
\affiliation{College of Physics and Technology, Guangxi Normal University, Guilin, Guangxi 541004, China}

\date{\today}

\begin{abstract}
The Bayesian reconstruction entropy is considered an alternative to the Shannon-Jaynes entropy, as it does not exhibit the asymptotic flatness characteristic of the Shannon-Jaynes entropy and obeys the scale invariance. It is commonly utilized in conjunction with the maximum entropy method to derive spectral functions from Euclidean time correlators produced by lattice QCD simulations. This study expands the application of the Bayesian reconstruction entropy to the reconstruction of spectral functions for Matsubara or imaginary-time Green's functions in quantum many-body physics. Furthermore, it extends the Bayesian reconstruction entropy to implement the positive-negative entropy algorithm, enabling the analytic continuations of matrix-valued Green's functions on an element-wise manner. Both the diagonal and off-diagonal components of the matrix-valued Green's functions are treated equally. Benchmark results for the analytic continuations of synthetic Green's functions indicate that the Bayesian reconstruction entropy, when combined with the preblur trick, demonstrates comparable performance to the Shannon-Jaynes entropy. Notably, it exhibits greater resilience to noises in the input data, particularly when the noise level is moderate.
\end{abstract}

\maketitle

\section{Introduction\label{sec:intro}}

In quantum many-body physics, single-particle Green's function holds significant importance~\cite{many_body_book,many_body_book_2016}. It is typically computed using numerical methods such as quantum Monte Carlo~\cite{RevModPhys.83.349,RevModPhys.73.33,gubernatis_kawashima_werner_2016} and functional renormalization group~\cite{RevModPhys.84.299,doi:10.1080/00018732.2013.862020}, or many-body perturbative techniques like random phase approximation~\cite{PhysRevLett.103.056401,annurev-physchem-040215-112308} and GW approximation~\cite{PhysRev.139.A796,Aryasetiawan_1998,RevModPhys.74.601,PhysRevB.91.235114}, within the framework of imaginary time or Matsubara frequency formulation. This quantity provides valuable insights into the dynamic response of the system. However, it cannot be directly compared to real-frequency experimental observables. Therefore, it is necessary to convert the calculated values to real frequencies, which presents the fundamental challenge of the analytic continuation problem. In theory, the retarded Green's function $G^{R}(\omega)$ and the spectral function $A(\omega)$ can be derived from the imaginary-time Green's function $G(\tau)$ or Matsubara Green's function $G(i\omega_n)$ by performing an inverse Laplace transformation~\cite{many_body_book,many_body_book_2016}. Due to the ill-conditioned nature of the transformation kernel, solving it directly is nearly impossible. In fact, the output for this inverse problem is highly sensitive to the input. Even minor fluctuations or noises in $G(\tau)$ or $G(i\omega_n)$ can result in nonsensical changes for $A(\omega)$ or $G^{R}(\omega)$, making analytic continuations extremely unstable.

In recent decades, numerous techniques have been developed to address the challenges associated with analytic continuation problems. These methods include Pad\'{e} approximation (PA)~\cite{PhysRevB.61.5147,PhysRevB.93.075104,Vidberg1977,PhysRevD.96.036002}, maximum entropy method (MaxEnt)~\cite{JARRELL1996133,PhysRevB.44.6011,PhysRevB.41.2380}, stochastic analytic continuation (SAC)~\cite{PhysRevB.57.10287,PhysRevE.94.063308,SHAO20231,beach2004,PhysRevE.81.056701,PhysRevB.76.035115,doi:10.1063/1.3185728,PhysRevB.102.035114,PhysRevB.101.085111,PhysRevB.78.174429}, stochastic optimization method (SOM)~\cite{PhysRevB.62.6317,PhysRevB.94.125149,PhysRevB.95.014102,KRIVENKO2019166,KRIVENKO2022108491}, stochastic pole expansion (SPX)~\cite{PhysRevB.108.235143,huang2023reconstructing}, sparse modeling (SpM)~\cite{PhysRevE.95.061302,PhysRevB.105.035139}, Nevanlinna analytical continuation (NAC)~\cite{PhysRevLett.126.056402,PhysRevB.104.165111}, causal projections~\cite{PhysRevB.107.075151}, Prony fits~\cite{zhang2023minimal,Ying2022,YING2022111549}, and machine learning assistant approaches~\cite{PhysRevLett.124.056401,PhysRevB.98.245101,PhysRevB.105.075112,PhysRevResearch.4.043082,Yao_2022,Arsenault_2017,PhysRevE.106.025312}, and so on. Each method has its own advantages and disadvantages. There is no doubt that the MaxEnt method is particularly popular due to its ability to maintain a balance between computational efficiency and accuracy~\cite{JARRELL1996133,PhysRevB.44.6011,PhysRevB.41.2380}. Several open source toolkits, such as \texttt{$\Omega$MAXENT}~\cite{PhysRevE.94.023303}, \texttt{QMEM}~\cite{PhysRevB.98.205102}, \texttt{MAXENT} by Kraberger \emph{et al.}~\cite{PhysRevB.96.155128}, \texttt{MAXENT} by Levy \emph{et al.}~\cite{LEVY2017149}, \texttt{ALF}~\cite{10.21468/SciPostPhysCodeb.1}, \texttt{ACFlow}~\cite{Huang:2022}, \texttt{eDMFT}~\cite{PhysRevB.81.195107}, and \texttt{ana\_cont}~\cite{KAUFMANN2023108519}, have been developed to implement this method.

The MaxEnt method is rooted in Bayesian inference~\cite{statistics_book}. Initially, the spectral function $A(\omega)$ is constrained to be non-negative and is interpreted as a probability distribution. Subsequently, the MaxEnt method endeavors to identify the most probable spectrum by minimizing a functional $Q[A]$, which comprises the misfit functional $L[A]$ and the entropic term $S[A]$ [see Eq.~(\ref{eq:qa})]. Here, $L[A]$ quantifies the difference between the input and reconstructed Green's function, while $S[A]$ is employed to regulate the spectrum~\cite{JARRELL1996133}. The MaxEnt method is well-established for the diagonal components of matrix-valued Green's functions, as the corresponding spectral functions are non-negative (for fermionic systems). However, the non-negativity of the spectral functions cannot be assured for the off-diagonal components of matrix-valued Green's functions. As a result, conventional MaxEnt method fails in this situation.

Currently, there is a growing emphasis on obtaining the full matrix representation of the Green's function (or the corresponding self-energy function) on the real axis~\cite{PhysRevLett.99.126402,PhysRevB.98.205102}. This is necessary for calculating lattice quantities of interest in Dyson's equation~\cite{PhysRevB.81.195107}. Therefore, there is a strong demand for a reliable method to perform analytic continuation of the entire Green's function matrix~\cite{PhysRevB.107.075151,zhang2023minimal,PhysRevB.104.165111}. To meet this requirement, several novel approaches have been suggested in recent years as enhancements to the conventional MaxEnt method. The three most significant advancements are as follows: (1) \emph{Auxiliary Green's function algorithm}. It is possible to create an auxiliary Green's function by combining the off-diagonal and diagonal elements of the matrix-valued Green's function to ensure the positivity of the auxiliary spectrum~\cite{Tomczak_2007,PhysRevB.90.041110,PhysRevB.92.060509,PhysRevB.95.121104,PhysRevLett.110.216405,yue2023maximum}. The analytic continuation of the auxiliary Green's function can then be carried out using the standard MaxEnt method. (2) \emph{Positive-negative entropy algorithm}. Kraberger \emph{et al.} proposed that the off-diagonal spectral functions can be seen as a subtraction of two artificial positive functions. They extended the entropic term to relax the non-negativity constraint presented in the standard MaxEnt method~\cite{PhysRevB.96.155128,KAUFMANN2023108519}. This enables simultaneous treatment of both the diagonal and off-diagonal elements, thereby restoring crucial constraints on the mathematical properties of the resulting spectral functions, including positive semi-definiteness and Hermiticity. (3) \emph{Maximum quantum entropy algorithm}. Recently, Sim and Han generalized the MaxEnt method by reformulating it with quantum relative entropy, maintaining the Bayesian probabilistic interpretation~\cite{statistics_book,JARRELL1996133}. The matrix-valued Green's function is directly continued as a single object without any further approximation or ad-hoc treatment~\cite{PhysRevB.98.205102}.

Although various algorithms have been proposed to enhance the usefulness of the MaxEnt method~\cite{PhysRevB.96.155128,PhysRevE.94.023303,PhysRevB.98.205102}, further improvements in algorithms and implementations are always beneficial. As mentioned above, the essence of the MaxEnt method lies in the definition of the entropic term $S[A]$. Typically, it takes the form of Shannon-Jaynes entropy (dubbed $S_{\text{SJ}}$) in the realm of quantum many-body physics~\cite{PhysRev.106.620}. However, alternative forms such as the Tikhonov regulator~\cite{tikhonov}, the positive-negative entropy~\cite{PhysRevB.96.155128}, and the quantum relative entropy~\cite{PhysRevB.98.205102} are also acceptable. It is worth noting that in the context of lattice QCD simulations, the extraction of spectral functions from Euclidean time correlators is of particular important. The MaxEnt method is also the primary tool for the analytic continuation of lattice QCD data~\cite{ASAKAWA2001459}. Burnier and Rothkopf introduced an enhanced dimensionless prior distribution, known as the Bayesian reconstruction entropy (dubbed $S_{\text{BR}}$)~\cite{PhysRevLett.111.182003,fphy.2022.1028995}, which demonstrates superior asymptotic behavior compared to $S_{\text{SJ}}$~\cite{PhysRev.106.620}. They discovered that by using $S_{\text{BR}}$ in conjunction with the MaxEnt method, a significant improvement in the reconstruction of spectral functions for Euclidean time correlators can be achieved. They latter extended $S_{\text{BR}}$ to support Bayesian inference of non-positive spectral functions in quantum field theory~\cite{PhysRevD.95.056016}. But to the best of our knowledge, the application of $S_{\text{BR}}$ has been limited to post-processing for lattice QCD data thus far. Hence, some questions naturally arise. How effective is $S_{\text{BR}}$ for solving analytic continuation problems in quantum many-body simulations? Is it truly superior to $S_{\text{SJ}}$? Keeping these questions in mind, the primary objective of this study is to broaden the application scope of $S_{\text{BR}}$. We at first verify whether $S_{\text{BR}}$ can handle Matsubara (or imaginary-time) Green's functions. Then, we generalize $S_{\text{BR}}$ to implement the positive-negative entropy algorithm~\cite{PhysRevB.96.155128} for analytic continuations of matrix-valued Green's functions. The simulated results indicate that $S_{\text{BR}}$ works quite well in transforming imaginary-time or Matsubara Green's function data to real frequency, no matter whether the Green's function is a matrix or not. We find that $S_{\text{BR}}$ has a tendency to sharpen the peaks in spectral functions, but this shortcoming can be largely overcame by the preblur trick~\cite{KAUFMANN2023108519,PhysRevB.96.155128}. Overall, the performance of $S_{\text{BR}}$ is comparable to that of $S_{\text{SJ}}$~\cite{PhysRev.106.620} for the examples involved.

The rest of this paper is organized as follows. Section~\ref{sec:method} introduces the spectral representation of the single-particle Green's function. Furthermore, it explains the fundamental formalisms of the MaxEnt method, the Bayesian reconstruction entropy, and the principle of the positive-negative entropy algorithm. Section~\ref{sec:results} is dedicated to various benchmark examples, including analytic continuation results for synthetic single-band Green's functions and multi-orbital matrix-valued Green's functions. The spectral functions obtained by $S_{\text{SJ}}$ and $S_{\text{BR}}$ are compared with the exact spectra. In Section~\ref{sec:discussions}, we further examine the preblur trick and the auxiliary Green's function algorithm for $S_{\text{SJ}}$ and $S_{\text{BR}}$. The robustness of $S_{\text{BR}}$ with respect to noisy Matsubara data is discussed and compared with that of $S_{\text{SJ}}$. A concise summary is presented in Section~\ref{sec:summary}. Finally, Appendix~\ref{sec:goodness} introduces the goodness-of-fit functional. The detailed mathematical derivations for $S_{\text{SJ}}$ and $S_{\text{BR}}$ are provided in Appendices~\ref{sec:sj} and \ref{sec:br}, respectively.

\section{Method\label{sec:method}}

\subsection{Spectral representation of single-particle Green's function}

It is well known that the single-particle imaginary-time Green's function $G(\tau)$ can be defined as follows:
\begin{equation}
G(\tau) = \langle \mathcal{T}_{\tau} d(\tau) d^{\dagger}(0)\rangle,
\end{equation}
where $\tau$ is imaginary time, $\mathcal{T}_{\tau}$ means the time-ordered operator, $d$ ($d^{\dagger}$) denotes the annihilation (creation) operator~\cite{many_body_book,many_body_book_2016}. Given $G(\tau)$, Matsubara Green's function $G(i\omega_n)$ can be constructed via direct Fourier transformation:
\begin{equation}
G(i\omega_n) = \int^{\beta}_0 d\tau~e^{-i\omega_n \tau} G(\tau),
\end{equation}
where $\beta$ is the inverse temperature of the system ($\beta \equiv 1/T$), and $\omega_n$ is the Matsubara frequency. Note that $\omega_n$ is equal to $(2n+1)\pi /\beta$ for fermions and $2n\pi / \beta$ for bosons ($n$ is an integer).

Let us assume that the spectral function of the single-particle Green's function is $A(\omega)$, then the spectral representations of $G(\tau)$ and $G(i\omega_n)$ read:
\begin{equation}
\label{eq:gtau}
G(\tau) = \int^{+\infty}_{-\infty} K(\tau,\omega) A(\omega) d\omega,
\end{equation}
and
\begin{equation}
\label{eq:giwn}
G(i\omega_n) = \int^{+\infty}_{-\infty} K(i\omega_n,\omega) A(\omega) d\omega,
\end{equation}
respectively. Here, $K(\tau,\omega)$ and $K(i\omega_n,\omega)$ are called the kernel functions. Their definitions are as follows:
\begin{equation}
\label{eq:ktau}
K(\tau,\omega) = \frac{e^{-\tau\omega}}{1 \pm e^{-\beta\omega}},
\end{equation}
and
\begin{equation}
\label{eq:kiwn}
K(i\omega_n,\omega) = \frac{1}{i\omega_n - \omega}.
\end{equation}
In the right-hand side of Eq.~(\ref{eq:ktau}), + is for fermionic correlators and - is for bosonic correlators~\cite{JARRELL1996133}. In the subsequent discussion, our focus will be on fermionic correlators.  

We observe that Eqs.~(\ref{eq:gtau}) and (\ref{eq:giwn}) are classified as Fredholm integral equations of the first kind. When $A(\omega)$ is given, it is relatively straightforward to compute the corresponding $G(\tau)$ and $G(i\omega_n)$ by numerical integration methods. However, the inverse problem of deducing $A(\omega)$ from $G(\tau)$ or $G(i\omega_n)$ is challenging due to the ill-conditioned nature of the kernel function $K$. To be more specific, the condition number of $K$ is very large because of the exponential decay of $K$ with $\omega$ and $\tau$. Consequently, direct inversion of $K$ becomes impractical from a numerical standpoint~\cite{PhysRevB.44.6011}. Furthermore, the $G(\tau)$ or $G(i\omega_n)$ data obtained from finite temperature quantum Monte Carlo simulations is not free from errors~\cite{RevModPhys.83.349,RevModPhys.73.33,gubernatis_kawashima_werner_2016}, further complicating the solution of Eqs.~(\ref{eq:gtau}) and (\ref{eq:giwn}).

\subsection{Maximum entropy method}

The cornerstone of the MaxEnt method is Bayes' theorem. Let us treat the input Green's function $G$ and the spectrum $A$ as two events. According to Bayes' theorem, the posterior probability $P[A|G]$ can be calculated by:
\begin{equation}
P[A|G] = \frac{P[G|A] P[A]}{P[G]},
\end{equation}
where $P[G|A]$ is the likelihood function, $P[A]$ is the prior probability, and $P[G]$ is the evidence~\cite{statistics_book}. $P[G|A]$ is assumed to be in direct proportion to $e^{-L[A]}$, where the misfit functional $L[A]$ reads:
\begin{equation}
\label{eq:loss}
L[A] = \frac{1}{2} \chi^2[A] =
\frac{1}{2} \sum^N_{i=1} \left( \frac{G_i - \tilde{G}_i[A]}{\sigma_i} \right)^2.
\end{equation}
Here, $N$ is the number of input data points, $\sigma$ denotes the standard deviation of $G$, $\tilde{G}[A]$ is the reconstructed Green's function via Eqs.~(\ref{eq:gtau}) and (\ref{eq:giwn}), and $\chi^2[A]$ is called the goodness-of-fit functional (see Appendix~\ref{sec:goodness} for more details). On the other hand, $P[A]$ is supposed to be in direct proportion to $e^{\alpha S[A]}$, where $\alpha$ is a regulation parameter and $S$ is entropy. Since the evidence $P[G]$ can be viewed as a normalization constant, it is ignored in what follows. Thus,
\begin{equation}
P[A|G] \propto e^{Q},
\end{equation}
where $Q$ is a functional of $A$~\cite{JARRELL1996133}. It is defined as follows:
\begin{equation}
\label{eq:qa}
Q[A] = \alpha S[A] - L[A] = \alpha S[A] - \frac{1}{2} \chi^2 [A].
\end{equation}
The basic idea of the MaxEnt method is to identify the optimal spectrum $\hat{A}$ that maximizes the posterior probability $P[A|G]$ (or equivalently $Q[A]$). In essence, our goal is to determine the most favorable $\hat{A}$ that satisfies the following equation:
\begin{equation}
\label{eq:opt}
\frac{\partial Q}{\partial A} \Big|_{A = \hat{A}} = 0.
\end{equation}
Eq.~(\ref{eq:opt}) can be easily solved by the standard Newton's algorithm~\cite{optimization_book}. In Appendices~\ref{sec:goodness}, \ref{sec:sj} and \ref{sec:br}, all terms in the right-hand side of Eq.~(\ref{eq:qa}) are elaborated in detail. We also explain how to solve Eq.~(\ref{eq:opt}) efficiently.

\subsection{Bayesian reconstruction entropy}

The entropy $S$ is also known as the Kullback-Leibler distance sometimes. In principle, there are multiple choices for it. Perhaps $S_{\text{SJ}}$ is the most frequently used in quantum many-body physics and condensed matter physics~\cite{PhysRev.106.620,JARRELL1996133}. It reads:
\begin{equation}
\label{eq:sje}
S_{\text{SJ}}[A] = \int d\omega
\left[
    A(\omega) - D(\omega) - A(\omega) \log \left(\frac{A(\omega)}{D(\omega)}\right)
\right].
\end{equation}
Here, $D(\omega)$ is called the default model, providing the essential features of spectra. Usually $D(\omega)$ is a constant or a Gaussian function, but it can be determined by making use of high-frequency behavior of input data as well~\cite{PhysRevE.94.023303}. If both $A(\omega)$ and $D(\omega)$ have the same normalization, Eq.~(\ref{eq:sje}) reduces to:
\begin{equation}
S_{\text{SJ}}[A] = -\int d\omega~A(\omega)
\log \left(\frac{A(\omega)}{D(\omega)}\right).
\end{equation}
The $S_{\text{BR}}$ introduced by Burnier and Rothkopf~\cite{PhysRevLett.111.182003,fphy.2022.1028995} is dominant in high-energy physics and particle physics. It reads:
\begin{equation}
\label{eq:bre}
S_{\text{BR}}[A] = \int d\omega
\left[
    1 - \frac{A(\omega)}{D(\omega)} + \log \left(\frac{A(\omega)}{D(\omega)}\right)
\right].
\end{equation}
Note that $S_{\text{SJ}}$ is constructed from four axioms~\cite{JARRELL1996133}. While for $S_{\text{BR}}$, two of these axioms are replaced~\cite{PhysRevLett.111.182003}. First of all, the scale invariance is incorporated in $S_{\text{BR}}$. It means that $S_{\text{BR}}$ only depends on the ratio between $A$ and $D$. The integrand in Eq.~(\ref{eq:bre}) is dimensionless, such that the choice of units for $A$ and $D$ will not change the result of the spectral reconstruction. Second, $S_{\text{BR}}$ favors choosing smooth spectrum. The spectra that deviate between two adjacent frequencies, $\omega_1$ and $\omega_2$, should be penalized.

\subsection{Positive-negative entropy}

The formulas discussed above are only correct for positive definite spectra with a finite norm, i.e.,
\begin{equation}
\int d\omega~A(\omega) > 0.
\end{equation}
However, the norm could be zero for off-diagonal elements of matrix-valued Green's functions due to the fermionic anti-commutation relation~\cite{many_body_book,PhysRevB.107.075151}:
\begin{equation}
\int d\omega~A(\omega) = 0.
\end{equation}
This suggests that the spectral function is not positive definite any more. The equations for $S_{\text{SJ}}$ and $S_{\text{BR}}$ [i.e., Eqs.~(\ref{eq:sje}) and (\ref{eq:bre})] should be adapted to this new circumstance. The positive-negative entropy algorithm proposed by Kraberger \emph{et al.} is a graceful solution to this problem~\cite{PhysRevB.96.155128,KAUFMANN2023108519}. They rewrite the off-diagonal spectral function $A(\omega)$ as the subtraction of two positive definite spectra:
\begin{equation}
A(\omega) = A^{+}(\omega) - A^{-}(\omega).
\end{equation}
Here $A^{+}(\omega)$ and $A^{-}(\omega)$ are independent, but have the same norm. Then the resulting entropy can be split into two parts:
\begin{equation}
S^{\pm}[A] = S[A^{+},A^{-}] = S[A^+] + S[A^-].
\end{equation}
$S^{\pm}[A]$ (or $S[A^{+},A^{-}]$) is called the positive-negative entropy, which was first used for the analysis of NMR spectra~\cite{LAUE1985418,Sibisi1990}. The expressions of positive-negative entropy for $S_{\text{SJ}}$ and $S_{\text{BR}}$ are as follows:
\begin{widetext}
\begin{equation}
\label{eq:sj_pm}
S^{\pm}_{\text{SJ}}[A] =
\int d\omega~
\left [
    \sqrt{A^2 + 4D^2} - 2D - A \log \left(\frac{\sqrt{A^2 + 4D^2} + A}{2D}\right)
\right],
\end{equation}
and
\begin{equation}
\label{eq:br_pm}
S^{\pm}_{\text{BR}}[A] =
\int d\omega~
\left[
    2 - \frac{\sqrt{A^2 + D^2} + D}{D} +
    \log \left(\frac{\sqrt{A^2 + D^2} + D}{2D}\right)
\right].
\end{equation}
\end{widetext}
The default models for $A^{+}$ and $A^{-}$ are $D^{+}$ and $D^{-}$, respectively. To derive Eqs.~(\ref{eq:sj_pm}) and (\ref{eq:br_pm}), $D^{+} = D^{-} = D$ is assumed. For a detailed derivation, please see Appendices~\ref{sec:sj} and \ref{sec:br}. Similar to $S_{\text{BR}}[A]$, $S^{\pm}_{\text{BR}}[A]$ also exhibits scale invariance. In other words, it depends on $A/D$ only. 

\subsection{Entropy density}

\begin{figure}[t]
\includegraphics[width=\columnwidth]{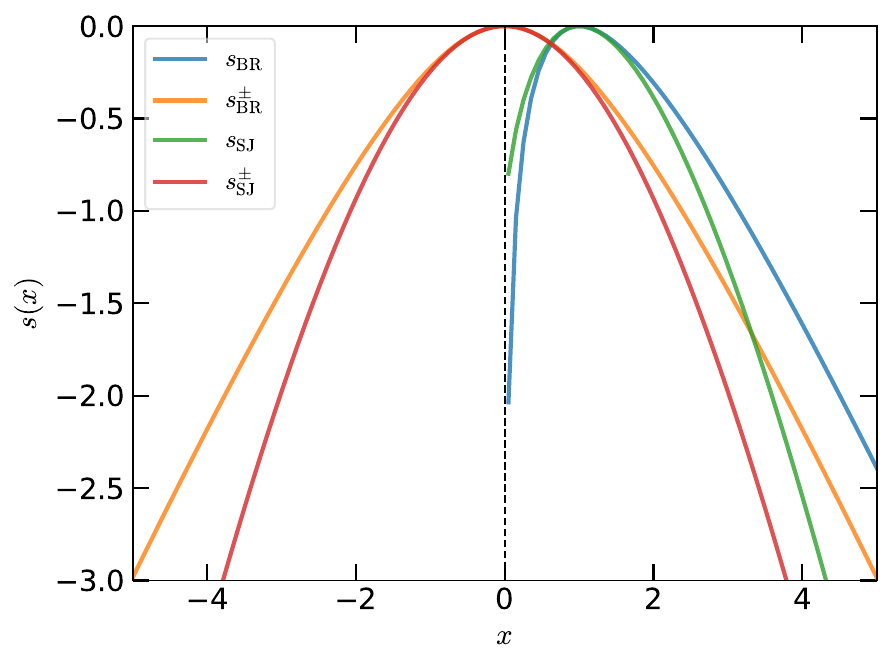}
\caption{Comparison of entropy densities for different entropic terms. The entropy $S$ is related to the entropy density $s$ via Eq.~(\ref{eq:density}). For the positive-negative entropy $S^{\pm}$, $D^{+} = D^{-} = D$ is assumed. \label{fig:entropy}}
\end{figure}

The integrand in the expression for entropy $S$ is referred as entropy density $s$, i.e.,
\begin{equation}
\label{eq:density}
S[A] = \int d\omega~s[A(\omega)].
\end{equation}
Next, we would like to make a detailed comparison about the entropy densities of different entropic terms. Supposed that $x = A/D$, the expressions for various entropy densities are collected as follows:
\begin{equation}
s_{\text{SJ}}(x) = x - 1 - x\log{x},
\end{equation}
\begin{equation}
s_{\text{BR}}(x) = 1 - x + \log{x},
\end{equation}
\begin{equation}
s^{\pm}_{\text{SJ}}(x) = \sqrt{x^2 + 4} - 2 - x \log{\frac{\sqrt{x^2 + 4}+x}{2}},
\end{equation}
\begin{equation}
s^{\pm}_{\text{BR}}(x) = 2 - (\sqrt{x^2 + 1} + 1) + \log{\frac{\sqrt{x^2+1}+1}{2}}.
\end{equation}
They are visualized in Figure~\ref{fig:entropy}.

Clearly, all the entropy densities are strictly convex and non-positive (i.e., $s \le 0$ and $s^{\pm} \le 0$). The ordinary entropy density $s(x)$ is just defined for positive $x$. $s_{\text{BR}}(x)$ becomes maximal at $x = 1$, and exhibits a similar quadratic behavior around its maximum as $s_{\text{SJ}}(x)$. In the case $x \to 0$ ($A \ll D$), $s_{\text{BR}}(x)$ is not suppressed, while $s_{\text{SJ}}(x) \to -1$. Thus, $s_{\text{BR}}(x)$ avoids the asymptotic flatness inherent in $s_{\text{SJ}}(x)$~\cite{PhysRevLett.111.182003}. The positive-negative entropy density $s^{\pm}(x)$ is valid for any $x$ ($x \in \mathbb{R}$). Both $s^{\pm}_{\text{BR}}(x)$ and $s^{\pm}_{\text{SJ}}(x)$ are even functions. They also exhibit quadratic behaviors around $x = 0$, and $s^{\pm}_{\text{BR}}(x) \ge s^{\pm}_{\text{SJ}}(x)$. In the limit of $\alpha \to \infty$, the goodness-of-fit functional $\chi^2[A]$ has negligible weight, and the entropic term $\alpha S[A]$ becomes dominant~\cite{PhysRevE.94.023303}. Thus, the maximization of $Q[A]$ (equivalently, the maximization of $S[A]$) yields $\hat{A} = D$ (at $x = 1$) for conventional entropy, in contrast to $\hat{A} = 0$ (at $x = 0$) for the positive-negative entropy~\cite{PhysRevB.96.155128}.

\section{Results\label{sec:results}}

\begin{figure*}[t]
\includegraphics[width=\textwidth]{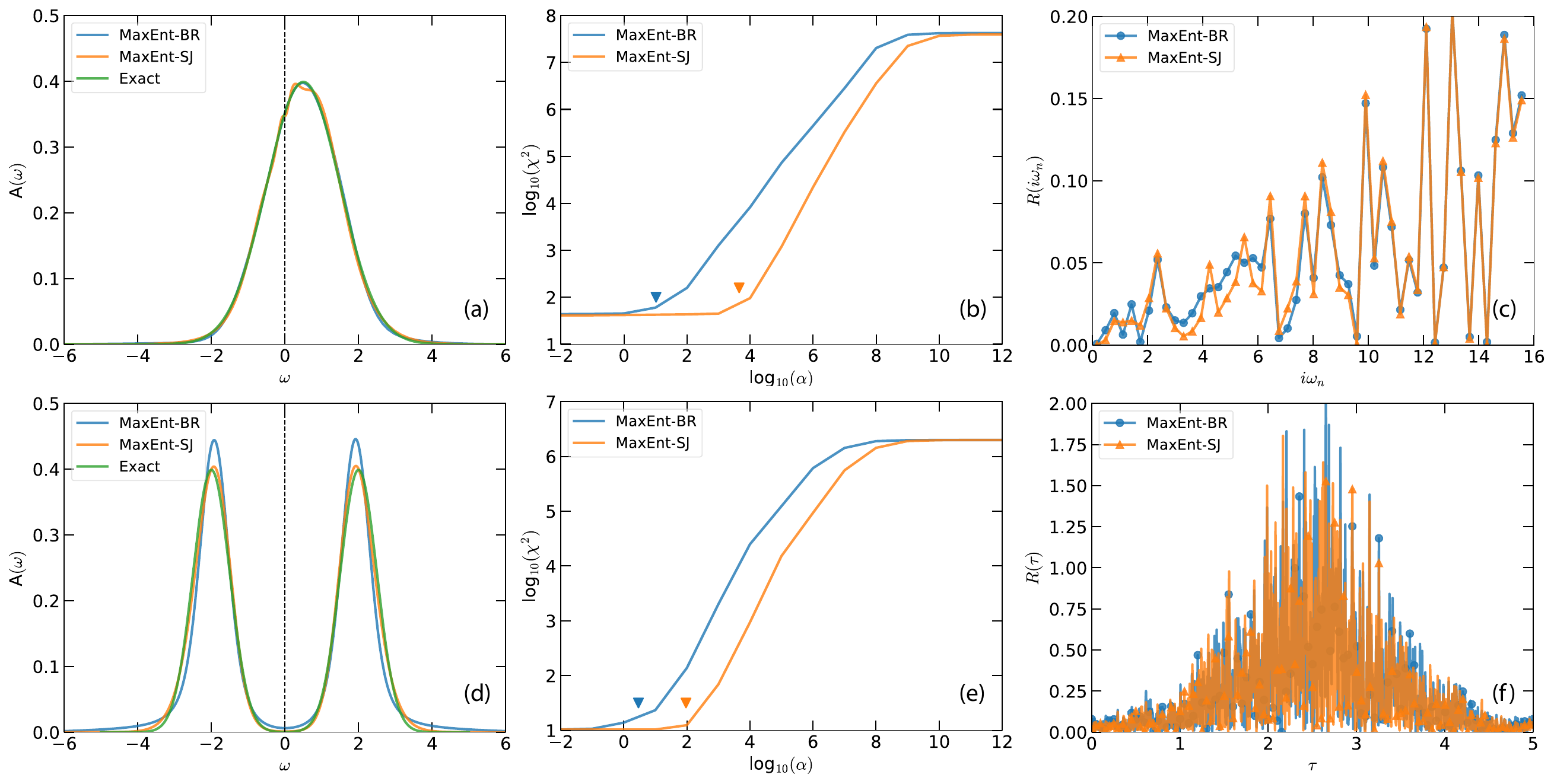}
\caption{Analytic continuations of single-band Green's functions. (a)-(c) For single off-centered peak case. (d)-(f) For two Gaussian peaks with a gap case. In panels (a) and (d), the calculated and exact spectra are compared. In panels (b) and (e), the $\log_{10}(\chi^2)-\log_{10}(\alpha)$ curves are plotted. The inverted triangle symbols indicate the optimal $\alpha$ parameters, which are determined by the $\chi^2$-kink algorithm~\cite{PhysRevE.94.023303}. The deviations of the reproduced Green's function from the raw ones are shown in panels (c) and (f). Note that in panel (c), $R$ is a function of Matsubara frequency. However, $R$ is $\tau$-dependent in panel (c). \label{fig:A06}}
\end{figure*}

\begin{figure*}[t]
\includegraphics[width=\textwidth]{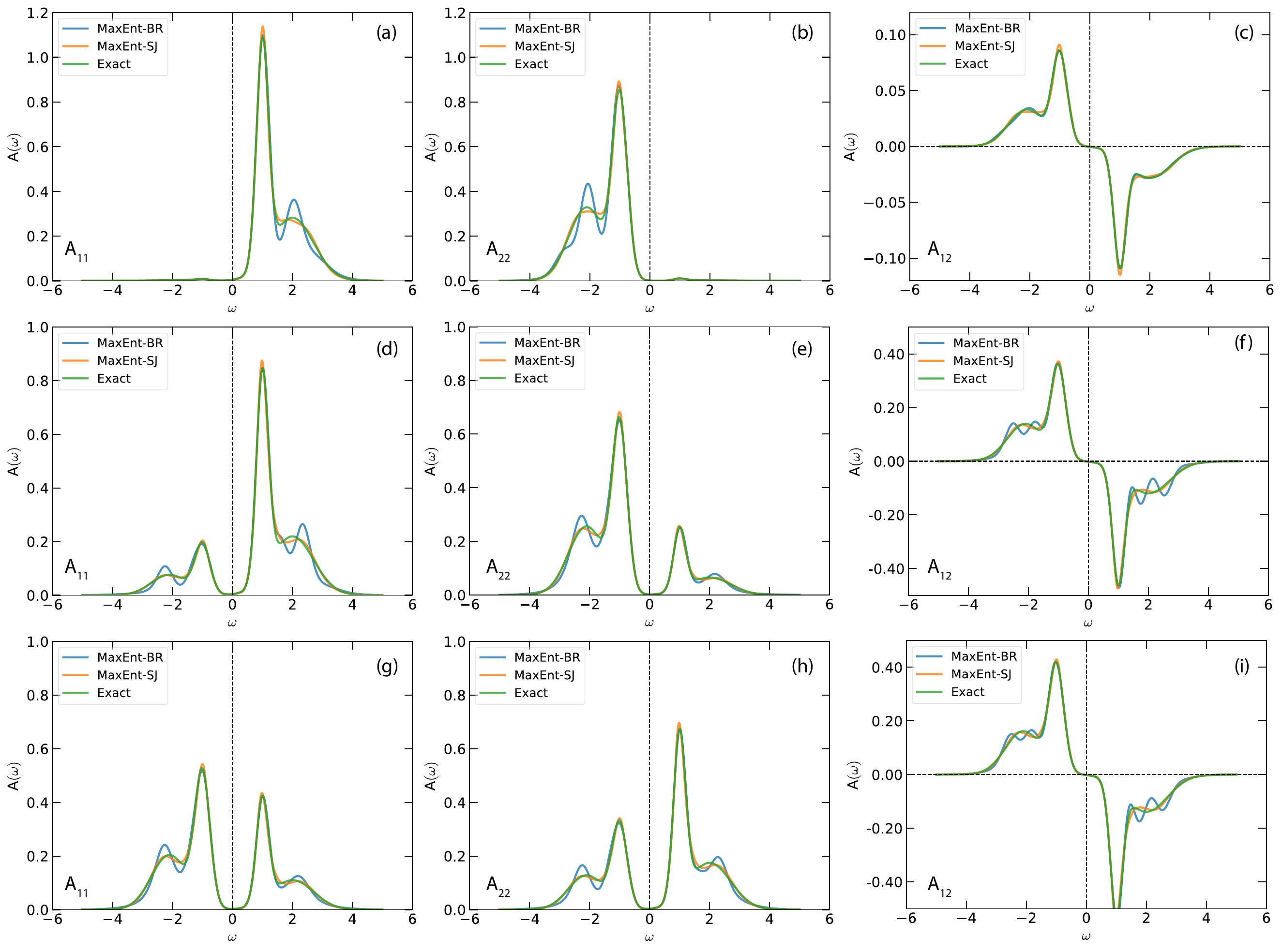}
\caption{Analytic continuations of two-band matrix-valued Green's functions. For analytic continuations for the diagonal elements of the Green's function, $S_{\text{SJ}}$ and $S_{\text{BR}}$ are used in the calculations. While for the off-diagonal elements of the Green's functions, $S^{\pm}_{\text{SJ}}$ and $S^{\pm}_{\text{BR}}$ are employed. In the MaxEnt-BR simulations, the preblur trick is always applied, and the blur parameter $b$ is fixed to 0.2. The default models for off-diagonal calculations are constructed by $A_{11}$ and $A_{22}$. See main text for more details. (a)-(c) $\theta = 0.1$; (d)-(f) $\theta = 0.5$; (g)-(i) $\theta = 0.9$. \label{fig:A08}}
\end{figure*}

\subsection{Computational setups}

When the conventional MaxEnt method is combined with $S_{\text{SJ}}$ (or $S^{\pm}_{\text{SJ}}$)~\cite{PhysRev.106.620}, it is called the MaxEnt-SJ method. On the other hand, a combination of the MaxEnt method with $S_{\text{BR}}$ (or $S^{\pm}_{\text{BR}}$)~\cite{PhysRevLett.111.182003,PhysRevD.95.056016,fphy.2022.1028995} is called the MaxEnt-BR method. We implemented both methods in the open source analytic continuation toolkit, namely \texttt{ACFlow}~\cite{Huang:2022}. All the benchmark calculations involved in this paper were done by this toolkit. The simulated results for the MaxEnt-SJ and MaxEnt-BR methods will be presented in this section. Next, we will explain the computational details.

We always start from an exact spectrum $A(\omega)$, which consists of one or more Gaussian-like peaks. Then $A(\omega)$ is used to construct the imaginary-time Green's function $G(\tau)$ via Eq.~(\ref{eq:gtau}), or the Matsubara Green's function $G(i\omega_n)$ via Eq.~(\ref{eq:giwn}). Here, for the sake of simplicity, the kernel function $K$ is assumed to be fermionic. Since the realistic Green's function data from finite temperature quantum Monte Carlo simulations is usually noisy, multiplicative Gaussian noise will be manually added to the clean $G(\tau)$ or $G(i\omega_n)$ to imitate this situation. The following formula is adopted:
\begin{equation}
\label{eq:noise}
G_{\text{noisy}} = G_{\text{clean}}[1 + \delta N_{\mathbb{C}}(0,1)],
\end{equation}
where $\delta$ denotes the noise level of the input data and $N_{\mathbb{C}}$ is the complex-valued normal Gaussian noise~\cite{PhysRevB.107.075151}. Unless stated otherwise, the standard deviation of $G$ is fixed to $10^{-4}$. The numbers of input data for $G(\tau)$ and $G(i\omega_n)$ are 1000 and 50, respectively. During the MaxEnt simulations, the $\chi^{2}$-kink algorithm~\cite{PhysRevE.94.023303} is used to determine the regulation parameter $\alpha$. The Bryan algorithm~\cite{Bryan1990} is also tested. It always gives similar results, which will not be presented in this paper. Once $S_{\text{BR}}$ and $S^{\pm}_{\text{BR}}$ are chosen, the preblur trick~\cite{KAUFMANN2023108519,PhysRevB.96.155128} is used to smooth the spectra. The blur parameter is adjusted case by case. Finally, the reconstructed spectrum is compared with the true solution. We also adopt the following quantity to quantify the deviation of the reconstructed Green's function from the input one:
\begin{equation}
R_i = 100 \left|\frac{\Delta G_i}{G_i}\right|
    = 100 \left|\frac{G_i - \tilde{G}_i[\hat{A}]}{G_i}\right|,
\end{equation}
where $\tilde{G}$ is evaluated by $\hat{A}$ via Eq.~(\ref{eq:gtau}) or Eq.~(\ref{eq:giwn}).

\subsection{Single-band Green's functions\label{subsec:one-band}}

At first, the analytic continuations of single-band Green's functions are examined. The exact spectral functions are constructed by a superposition of some Gaussian peaks. We consider two representative spectra. (i) Single off-centered peak. The spectral function is:
\begin{equation}
\label{eq:gauss1}
A(\omega) = \exp\left[-\frac{(\omega - \epsilon) ^ 2}{2\Gamma ^ 2}\right],
\end{equation}
where $\epsilon = 0.5$, $\Gamma = 1.0$, $\delta = 10^{-4}$, $\sigma = 10^{-4}$, and $\beta = 20.0$. The input data is Matsubara Green's function, which contains 50 Matsubara frequencies ($N = 50$). The blur parameter for the MaxEnt-BR method is 0.45. (ii) Two Gaussian peaks with a gap. The spectral function is:
\begin{eqnarray}
\label{eq:gauss2}
A(\omega) & = & \frac{f_1}{\sqrt{2\pi} \Gamma_1}
            \exp\left[-\frac{(\omega - \epsilon_1) ^ 2}{2\Gamma_1 ^ 2}\right]
            \nonumber \\
          & + & \frac{f_2}{\sqrt{2\pi} \Gamma_2}
            \exp\left[-\frac{(\omega - \epsilon_2) ^ 2}{2\Gamma_2 ^ 2}\right]
            ,
\end{eqnarray}
where $f_1 = f_2 = 1.0$, $\epsilon_1 = -\epsilon_2 = 2.0$, $\Gamma_1 = \Gamma_2 = 0.5$, $\delta = 10^{-4}$, $\sigma = 10^{-3}$, and $\beta = 5.0$. The synthetic data is imaginary-time Green's function, which contains 1000 imaginary time slices ($N = 1000$). The blur parameter for the MaxEnt-BR method is 0.30.

The analytic continuation results are shown in Fig.~\ref{fig:A06}. For case (i), it is clear that the exact spectrum and Matsubara Green's function are well reproduced by both the MaxEnt-SJ method and the MaxEnt-BR method [see Fig.~\ref{fig:A06}(a) and (c)]. For case (ii), the MaxEnt-SJ method works quite well. The MaxEnt-BR method can resolve the gap and the positions of the two peaks accurately. However, it slightly overestimates their heights. We also observe that the difference between $G(\tau)$ and $\tilde{G}(\tau)$ becomes relatively apparent in the vicinity of $\tau = \beta/2$ [see Fig.~\ref{fig:A06}(d) and (f)]. Figure~\ref{fig:A06}(b) and (e) plot $\log_{10} (\chi^2)$ as a function of $\log_{10} (\alpha)$. The plots can be split into three distinct regions~\cite{PhysRevE.94.023303}: (i) \emph{Default model region} ($\alpha \to \infty$). $\chi^2$ is relatively flat and goes to its global maximum. The entropic term $\alpha S$ is much larger than $\chi^2$. The obtained spectrum $A(\omega)$ resembles the default model $D(\omega)$. (ii) \emph{Noise-fitting region} ($\alpha \to 0$). $\chi^2$ approaches its global minimum, but it is larger than $\alpha S$. In this region, the calculated spectrum $A(\omega)$ tends to fit the noise in $G(\tau)$ or $G(i\omega_n)$. (iii) \emph{Information-fitting region}. $\chi^2$ increases with the increment of $\alpha$. It is comparable with $\alpha S$. The optimal $\alpha$ parameter is located at the crossover between the noise-fitting region and the information-fitting region. For $S_{\text{BR}}$ and $S_{\text{SJ}}$, their $\chi^{2}(\alpha)$ curves almost overlap at the default model region and the noise-fitting region, but differ at the information-fitting region. It indicates that the optimal $\alpha$ parameters for $S_{\text{SJ}}$ are larger than those for $S_{\text{BR}}$.

\subsection{Matrix-valued Green's functions\label{subsec:two-band}}

Next, the analytic continuations of matrix-valued Green's functions are examined. We consider a two-band model. The initial spectral function is a diagonal matrix:
\begin{equation}
\mathbf{A}'(\omega) =
\left[
\begin{array}{cc}
    A'_{11}(\omega) & 0 \\
    0 & A'_{22}(\omega) \\
\end{array}
\right].
\end{equation}
Here $A'_{11}(\omega)$ and $A'_{22}(\omega)$ are constructed by using Eq.~(\ref{eq:gauss2}). For $A'_{11}(\omega)$, the parameters are: $f_1 = f_2 = 0.5$, $\epsilon_1 = 1.0$, $\epsilon_2 = 2.0$, $\Gamma_1 = 0.20$, and $\Gamma_2 = 0.70$. For $A'_{22}(\omega)$, the parameters are: $f_1 = f_2 = 0.5$, $\epsilon_1 = -1.0$, $\epsilon_2 = -2.1$, $\Gamma_1 = 0.25$, and $\Gamma_2 = 0.60$. The true spectral function is a general matrix with non-zero off-diagonal components. It reads:
\begin{equation}
\mathbf{A}(\omega) =
\left[
\begin{array}{cc}
    A_{11}(\omega) & A_{12}(\omega) \\
    A_{21}(\omega) & A_{22}(\omega) \\
\end{array}
\right].
\end{equation}
The diagonal matrix $\mathbf{A}'(\omega)$ is rotated by a rotation matrix $\mathbf{R}$ to generate $\mathbf{A}(\omega)$. The rotation matrix $\mathbf{R}$ is defined as follows:
\begin{equation}
\label{eq:rotation}
\mathbf{R} =
\left[
\begin{array}{cc}
    \cos \theta & \sin \theta \\
    -\sin \theta & \cos \theta \\
\end{array}
\right],
\end{equation}
where $\theta$ denotes the rotation angle. In the present work, we consider three different rotation angles. They are: (i) $\theta = 0.1$, (ii) $\theta = 0.5$, and (iii) $\theta = 0.9$. Then $\mathbf{A}(\omega)$ is used to construct the matrix-valued Green's function $\mathbf{G}(i\omega_n)$ by using Eq.~(\ref{eq:giwn}). The essential parameters are: $\delta = 0.0$, $\sigma = 10^{-4}$, and $\beta = 40.0$. The input data contains 50 Matsubara frequencies. For analytic continuations of the diagonal elements of $\mathbf{G}(i\omega_n)$, the default models are Gaussian-like. However, for the off-diagonal elements, the default models are quite different. They are evaluated by $D_{12}(\omega) = \sqrt{A_{11}(\omega)A_{22}(\omega) }$, where $A_{11}(\omega)$ and $A_{22}(\omega)$ are the calculated spectral functions for diagonal elements~\cite{PhysRevB.96.155128,KAUFMANN2023108519}. This means that we have to carry out analytic continuations for the diagonal elements at first, and then use the diagonal spectra to prepare the default models for the off-diagonal elements. For the MaxEnt-BR method, the preblur algorithm is used to smooth the spectra. The blur parameter is fixed to 0.2.

The simulated results are illustrated in Fig.~\ref{fig:A08}. For the diagonal spectral spectral functions ($A_{11}$ and $A_{22}$), both the MaxEnt-SJ and MaxEnt-BR methods can accurately resolve the peaks that are close to the Fermi level. However, for the high-energy peaks, the MaxEnt-BR method tends to overestimate their heights and yield sharper peaks [see the peaks around $\omega = 2.0$ in Fig.~\ref{fig:A08}(a), (d), and (g)]. For the off-diagonal spectral functions, only $A_{12}$ is shown, since $A_{21}$ is equivalent to $A_{12}$. We observe that the major features of $A_{12}$ are well captured by both the MaxEnt-SJ and MaxEnt-BR methods. Undoubtedly, the MaxEnt-SJ method works quite well in all cases. The MaxEnt-BR method exhibits good performance when the rotation angle is small [$\theta = 0.1$, see Fig.~\ref{fig:A08}(c)]. But some wiggles emerge around $\omega = \pm 2.0$ when the rotation angle is large [$\theta = 0.5$ and $\theta = 0.9$, see Fig.~\ref{fig:A08}(f) and (i)]. We test more rotation angles ($0.0 < \theta < 2.0$). It seems that these wiggles won't be enhanced unless the blur parameter $b$ is decreased.

\section{Discussions\label{sec:discussions}}

\begin{figure}[t]
\includegraphics[width=\columnwidth]{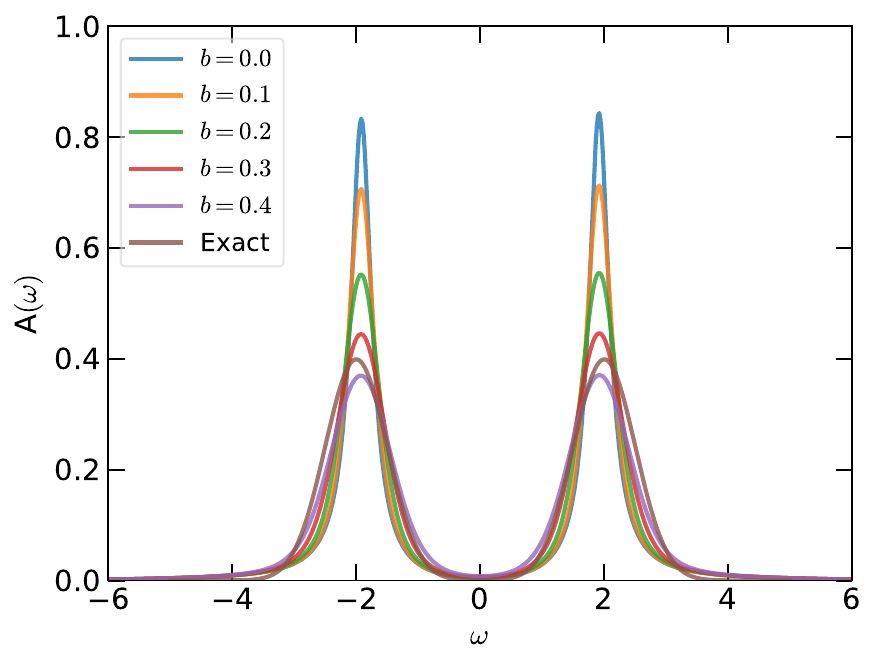}
\caption{Test of the preblur trick. A single-band Green's function is treated. The exact spectral function is generated by using Eq.~(\ref{eq:gauss2}). The model parameters are the same as those used in Sec.~\ref{subsec:one-band}. Only the analytic continuation results obtained by the MaxEnt-BR method are shown. \label{fig:blur1}}
\end{figure}

\begin{figure}[t]
\includegraphics[width=\columnwidth]{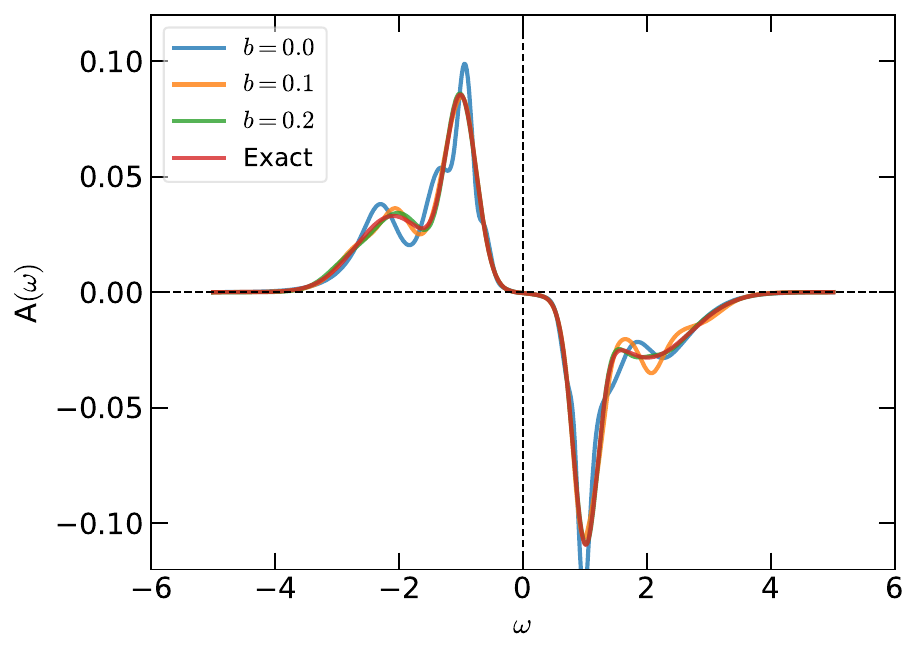}
\caption{Test of the preblur trick. A matrix-valued Green's function is treated via the positive-negative entropy approach. The initial spectral function is generated by using Eq.~(\ref{eq:gauss2}). The model parameters are the same as those used in Sec.~\ref{subsec:two-band}. The rotation angle $\theta$ is 0.1. Only the off-diagonal spectral functions ($A_{12}$) obtained by the MaxEnt-BR method are shown. \label{fig:blur2}}
\end{figure}

\begin{figure*}[t]
\includegraphics[width=\textwidth]{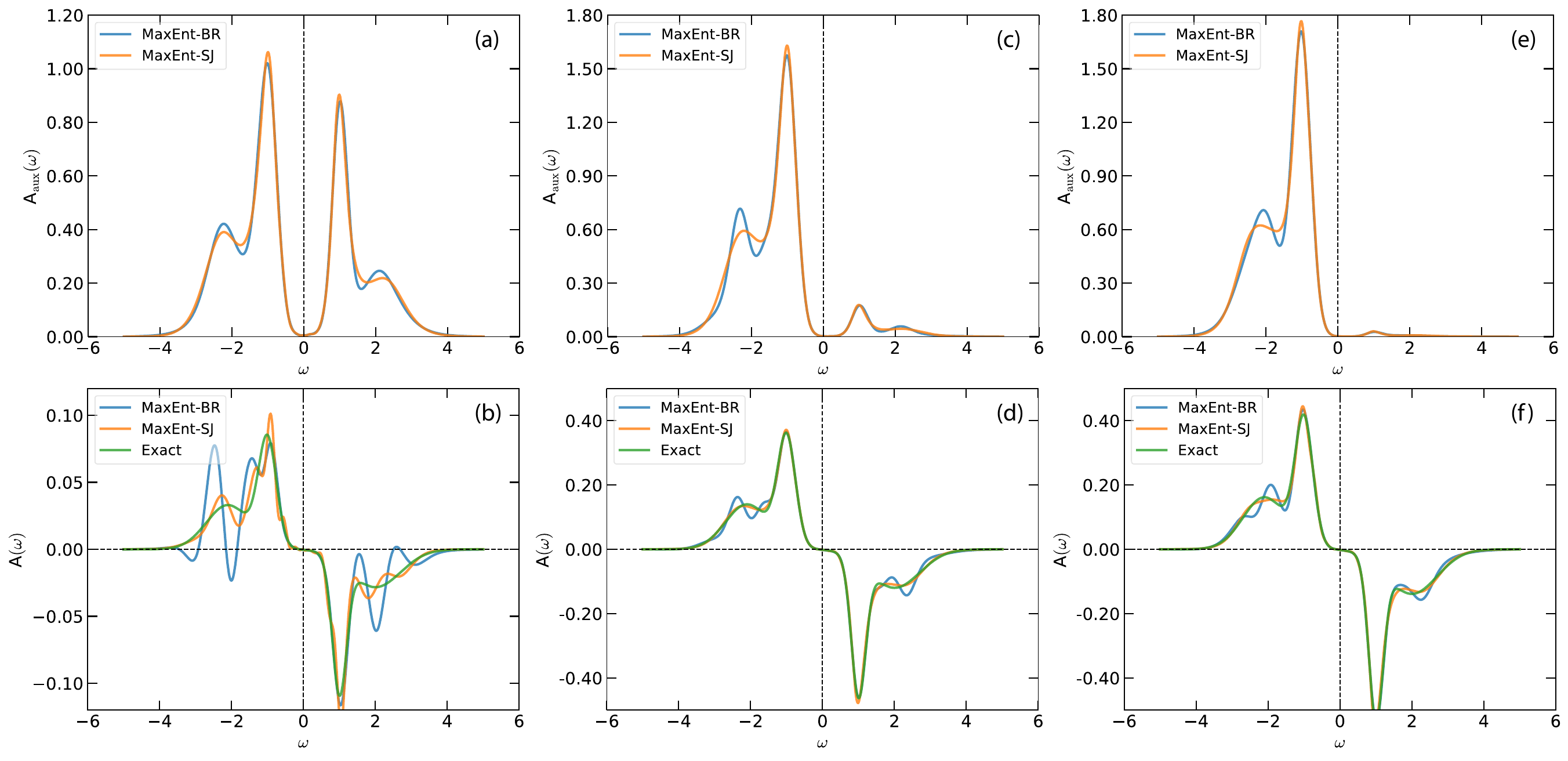}
\caption{Analytic continuations of matrix-valued Green's functions via the auxiliary Green's function approach. The model parameters are the same as those used in Sec.~\ref{subsec:two-band}. For the MaxEnt-BR simulations, the preblur trick is always applied, and the blur parameter $b$ is fixed to 0.2. Only the results for the off-diagonal elements are presented ($A_{12}$). The upper panels show the spectral functions for the auxiliary Green's functions [see Eq.~(\ref{eq:gaux})]. The lower panels show the off-diagonal spectral functions evaluated by Eq.~(\ref{eq:Aaux}). (a)-(b) $\theta = 0.1$; (c)-(d) $\theta = 0.5$; (e)-(f) $\theta = 0.9$. \label{fig:A08aux}}
\end{figure*}

\begin{figure*}[t]
\includegraphics[width=\textwidth]{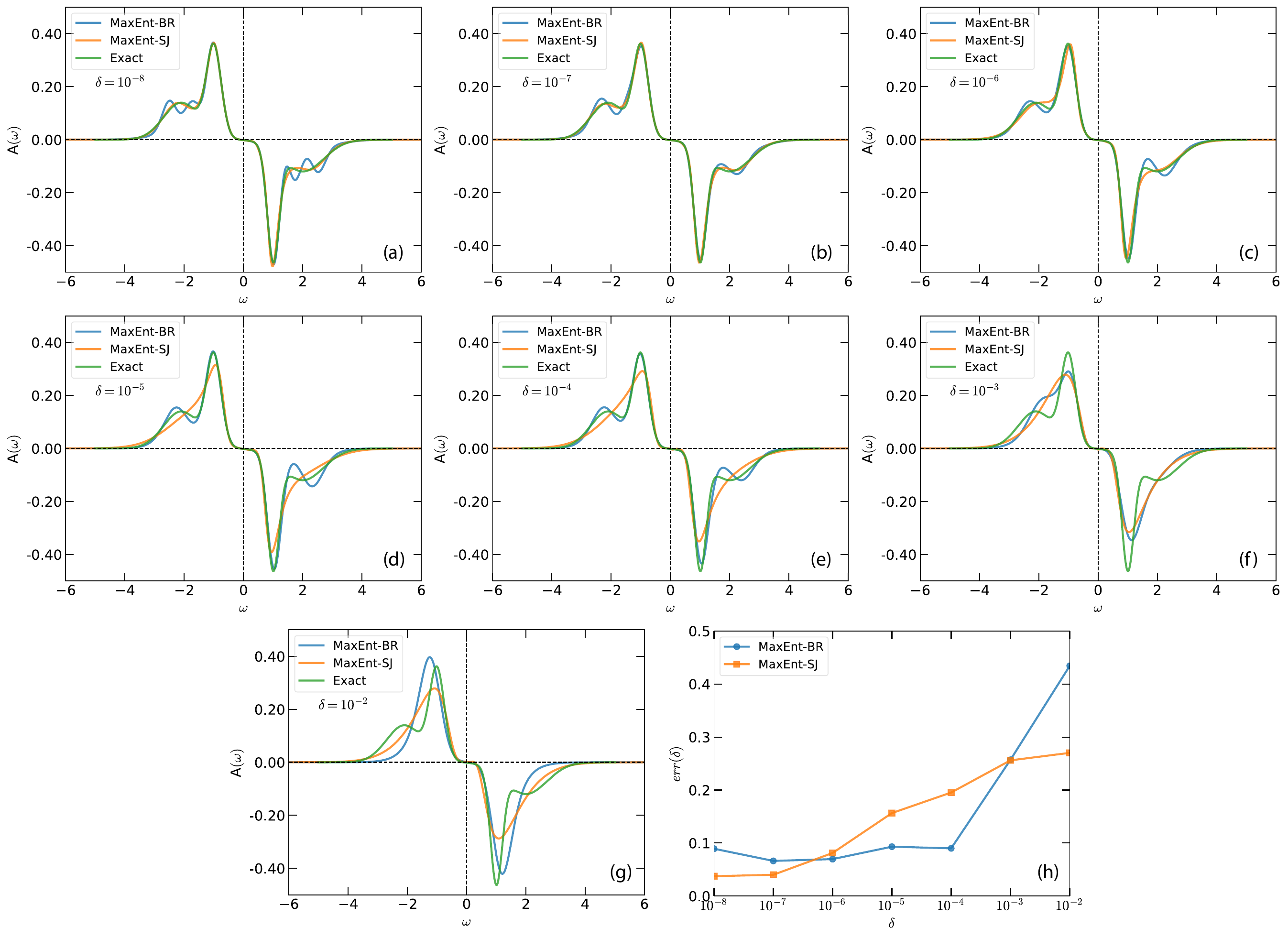}
\caption{Robustness of the MaxEnt method with respect to the noisy Matsubara data. Here we treat the off-diagonal elements ($G_{12}$) of the two-band matrix-valued Green's functions only via the positive-negative entropy method. The model parameters are the same as those used in Sec.~\ref{subsec:two-band}. The rotation angle $\theta$ is 0.5. The blur parameter for the MaxEnt-BR method is fixed to 0.2. (a)-(g) Off-diagonal spectral functions ($A_{12}$) extracted from noisy Matsubara data. $\delta$ denotes the noise level of the input data. It varies from $10^{-8}$ (tiny noise) to $10^{-2}$ (large noise). (h) Integrated real axis error as a function of $\delta$. It measures the deviations from the exact spectral function, see Eq.~(\ref{eq:err}). \label{fig:AN}}
\end{figure*}

In previous section, analytic continuations for single-band Green's functions and matrix-valued Green's functions by using the MaxEnt-BR method have been demonstrated. However, there are still some important issues that need to be clarified. In this section, we would like to further discuss the preblur algorithm, the auxiliary Green's algorithm, and the noise tolerance for the MaxEnt-BR method.

\subsection{Effect of preblur}

The benchmark results shown in Section~\ref{sec:results} imply that the MaxEnt-BR method has a tendency to generate sharp peaks or small fluctuations in the high-energy regions of the spectra. The preblur algorithm is helpful to alleviate this phenomenon. In the context of analytic continuation, the preblur algorithm was introduced by Kraberger \emph{et al.}~\cite{PhysRevB.96.155128}. The kernel function $K(i\omega_n,\omega)$ [see Eq.~(\ref{eq:kiwn})] is ``blurred'' by using the following expression:
\begin{equation}
K_{b}(i\omega_n,\omega) = \int^{\infty}_{-\infty} d\omega'~
    K(i\omega_n,\omega') g_b(\omega - \omega').
\end{equation}
Here, $K_{b}(i\omega_n,\omega)$ is the blurred kernel, which is then used in Eq.~(\ref{eq:loss}) to evaluate the $\chi^2$-term. $g_b(\omega)$ is a Gaussian function:
\begin{equation}
g_b(\omega) = \frac{\exp{(-\omega^2 / 2b^2)}}{\sqrt{2\pi} b},
\end{equation}
where $b$ is the blur parameter.

In Figures~\ref{fig:blur1} and \ref{fig:blur2}, we analyze the effects of the preblur trick for two typical scenarios: (i) Positive spectral function with two separate Gaussian peaks. (ii) Complicated spectral function for off-diagonal Green's function. Note that the model parameters for generating the exact spectra are taken from Sections~\ref{subsec:one-band} and \ref{subsec:two-band}, respectively. It is evident that the MaxEnt-BR method without the preblur trick ($b = 0$) has trouble resolving the spectra accurately. It usually favors sharp peaks (see Fig.~\ref{fig:blur1}). Sometimes it may lead to undesirable artifacts, such as the side peaks around $\omega = \pm 1.5$ in Fig.~\ref{fig:blur2}. The preblur algorithm can remedy this problem to some extent. The major peaks are smoothed, and the artificial side peaks are suppressed upon increasing $b$. But it's not the case that bigger is always better. There is an optimal $b$. For case (i), the optimal $b$ is 0.3. A larger $b$ ($b > 0.3$) will destroy the gap, inducing a metallic state. For case (ii), the optimal $b$ is 0.2. If $b$ is further increased, it is difficult to obtain a stable solution. Here, we should emphasize that these unphysical features seen in the spectra are not unique for the MaxEnt-BR method. Actually, a similar tendency was already observed in the early applications of the MaxEnt-SJ method in image processing tasks. In order to address this problem, Skilling suggested that the spectrum can be expressed as a Gaussian convolution: $A = g_b \star h$, where $h$ is a ``hidden'' function~\cite{10.1093/oso/9780198539414.003.0002}. Then the entropy is evaluated from $h(\omega)$, instead of $A(\omega)$. In fact, Skilling's approach is equivalent to the preblur trick.

\subsection{Auxiliary Green's functions}

As stated before, in addition to the positive-negative entropy method, the auxiliary Green's algorithm can be used to continue the off-diagonal elements of the Green's functions~\cite{Tomczak_2007,PhysRevB.90.041110,PhysRevB.92.060509,PhysRevB.95.121104,PhysRevLett.110.216405,yue2023maximum}. Its idea is quite simple. At first, an auxiliary Green's function is constructed as follows:
\begin{equation}
\label{eq:gaux}
G_{\text{aux}}(i\omega_n) = G_{11}(i\omega_n) + G_{22}(i\omega_n) + 2G_{12}(i\omega_n).
\end{equation}
Supposed that $A_{\text{aux}}(\omega)$ is the corresponding spectral function for $G_{\text{aux}}(i\omega_n)$. It is easy to prove that $A_{\text{aux}}(\omega)$ is positive semi-definiteness. So, it is safe to perform analytic continuation for $G_{\text{aux}}(i\omega_n)$ by using the traditional MaxEnt-SJ or MaxEnt-BR method. Next, the off-diagonal spectral function $A_{12}(\omega)$ can be evaluated as follows:
\begin{equation}
\label{eq:Aaux}
A_{12}(\omega) = \frac{A_{\text{aux}}(\omega) - A_{11}(\omega) - A_{22}(\omega)}{2}.
\end{equation}
This algorithm is not restricted to the particle-hole symmetric case, as assumed in Ref.~\cite{PhysRevB.90.041110}. Here, we employ the two-band's example presented in Section~\ref{subsec:two-band} again to test the combination of the auxiliary Green's function algorithm and the MaxEnt-BR method (and the MaxEnt-SJ method). The model and computational parameters are kept, and the analytic continuation results are plotted in Fig.~\ref{fig:A08aux}.

Three rotation angles are considered in the simulations. We find the auxiliary Green's function algorithm is numerically unstable. (i) $\theta = 0.1$. When the rotation angle is small, the amplitude (absolute value) of $A_{12}(\omega)$ is small. Both the MaxEnt-SJ method and the MaxEnt-BR method can resolve the major peaks around $\omega = \pm 1.0$. But they will produce apparent oscillations at higher energies. Increasing the $b$ parameter further could lead to wrong peaks. It seems that these unphysical features are likely due to the superposition of errors in $A_{\text{aux}}(\omega)$, $A_{11}(\omega)$, and $A_{22}(\omega)$. The performance of the MaxEnt-BR method is worse than that of the MaxEnt-SJ method. (ii) $\theta = 0.5$ and $\theta = 0.9$. When the rotation angle is moderate or large, the spectra obtained by the MaxEnt-SJ method are well consistent with the exact solutions. By using the MaxEnt-BR method, though the fluctuations in high-energy regions are greatly suppressed, small deviations from the exact spectra still exist. Overall, the auxiliary Green's function algorithm is inferior to the positive-negative entropy method in the examples studied. Especially when the rotation angle is small, the auxiliary Green's function algorithm usually fails, irrespective of which form of entropy is adopted.

\subsection{Robustness with respect to noisy input data}

Analytic continuation is commonly used on noisy Monte Carlo data. The precision of the Monte Carlo data, which strongly depends on the sampling algorithm and the estimator used, is rarely better than $10^{-5}$~\cite{RevModPhys.83.349,RevModPhys.73.33,gubernatis_kawashima_werner_2016}. The distributions of inherent errors in the Monte Carlo data are often Gaussian. This poses a strict requirement for the noise tolerance of the analytic continuation methods. Previous studies have demonstrated that the MaxEnt-SJ method is robust to noise. Here, we would like to examine the robustness of the MaxEnt-BR method with respect to noisy Matsubara data. We reconsider the analytic continuation of the off-diagonal elements of matrix-valued Green's functions. As mentioned above, the synthetic Matsubara data for this case is assumed to be noiseless ($\delta = 0.0$). Now the noise is manually added by Eq.~(\ref{eq:noise}) and the noise level $\delta$ is changed from $10^{-8}$ to $10^{-2}$. The other computational parameters are the same as those used in Section~\ref{subsec:two-band}. In order to estimate the sensitivity of the analytic continuation method to the noisy data, a new quantity, namely integrated real axis error, is introduced:
\begin{equation}
\label{eq:err}
err(\delta) = \int d\omega~\left|A(\omega) - \hat{A}_{\delta}(\omega)\right|.
\end{equation}
Here, $\hat{A}_{\delta}(\omega)$ means the reconstructed (optimal) spectral function under the given noise level $\delta$.

In Figure~\ref{fig:AN}(a)-(g), the convergence of the spectral functions by the MaxEnt-BR method for simulated Gaussian errors with varying magnitude is shown. When $\delta = 10^{-2}$, the main peaks at $\omega \approx \pm 1.0$ are roughly reproduced, while the side peaks at $\omega \approx 2.0$ are completely smeared out. When $\delta = 10^{-3}$, the major characteristics of the off-diagonal spectrum are successfully captured. As the simulated errors decrease further ($\delta < 10^{-3}$), the MaxEnt-BR method rapidly converges to the exact solution. For comparison, the results for the MaxEnt-SJ method are also presented. It is evident that the MaxEnt-SJ method is less robust than the MaxEnt-BR method when the noise level is moderate. It fails to resolve the satellite peaks around $\omega = \pm 2.0$. Only when $\delta \le 10^{-6}$, the MaxEnt-SJ method can recover the exact spectrum. Figure~\ref{fig:AN}(h) exhibits the integrated error $err(\delta)$. When $10^{-6} < \delta < 10^{-3}$, the MaxEnt-BR method exhibits better robustness with respect to noise than the MaxEnt-SJ method.      
 
\section{Concluding remarks\label{sec:summary}}

In summary, we extend the application scope of $S_{\text{BR}}$ to analytic continuations of imaginary-time Green's functions and Matsubara Green's functions in quantum many-body physics and condense matter physics. It is further generalized to the form of positive-negative entropy to support analytic continuation of matrix-valued Green's function, in which the positive semi-definiteness of the spectral function is broken. We demonstrate that the MaxEnt-BR method, in conjunction with the preblur algorithm and the positive-negative entropy algorithm, is capable of capturing the primary features of the diagonal and off-diagonal spectral functions, even in the presence of moderate levels of noise. Overall, its performance is on par with that of the MaxEnt-SJ method in the examples studied. Possible applications of the MaxEnt-BR method in the future include analytic continuations of anomalous Green's functions and self-energy functions~\cite{yue2023maximum,PhysRevB.90.041110}, bosonic response functions (such as optical conductivity and spin susceptibility)~\cite{PhysRevB.82.165125,PhysRevB.94.245140}, and frequency-dependent transport coefficients with non-positive spectral weight~\cite{PhysRevB.95.121104,PhysRevB.92.060509}, etc. Further investigations are highly desirable.    

Finally, the MaxEnt-BR method, together with the MaxEnt-SJ method, has been integrated into the open source software package \texttt{ACFlow}~\cite{Huang:2022}, which may be useful for the analytic continuation community.

\begin{acknowledgments}
This work is supported by the National Natural Science Foundation of China (No.~12274380 and No.~11934020), and the Central Government Guidance Funds for Local Scientific and Technological Development (No.~GUIKE~ZY22096024).
\end{acknowledgments}

\appendix

\section{Goodness-of-fit functional\label{sec:goodness}}

Given the spectral function $A(\omega)$, the Matsubara Green's function $\tilde{G}[A]$ could be reconstructed by using the following equation (the $\tilde{~}$ symbol is used to distinguish the reconstructed Green's function from the input Green's function):
\begin{equation}
\label{eq:KA_orig}
\tilde{G}_n[A] = \int d\omega~K(i\omega_n, \omega) A(\omega),
\end{equation}
where $K(i\omega_n,\omega)$ denotes the kernel function, and $n$ is the index for Matsubara frequency. For the sake of simplicity, Eq.~(\ref{eq:KA_orig}) is reformulated into its discretization form:
\begin{equation}
\label{eq:KA}
\tilde{G}_n[A] = \sum_m K_{nm} A_m \Delta_m,
\end{equation}
where $K_{nm} \equiv K(i\omega_n, \omega_m)$, $A_i  \equiv A(\omega_i)$, and $\Delta_i$ means the weight of the mesh at real axis.

The goodness-of-fit functional $\chi^{2}[A]$ measures the distance between the input Green's function $G$ and the reconstructed Green's function $\tilde{G}[A]$. Its expression is as follows:
\begin{equation}
\label{eq:chi2}
\chi^2 [A] = \sum^{N}_{n = 1} \frac{(G_n - \tilde{G}_n[A])^2 }{\sigma^2_n}
\end{equation}
Here we just assume that there are $N$ data points, and $\sigma_n$ denotes the standard derivative (i.e., the error bar) of $G_n$. Substituting Eq.~(\ref{eq:KA}) into Eq.~(\ref{eq:chi2}), then we arrive:
\begin{equation}
\chi^2 [A] = \sum^{N}_{n = 1} \frac{(G_n - \sum_m K_{nm} A_m \Delta_m)^2 }{\sigma^2_n}.
\end{equation}
The first derivative of $\chi^2 [A]$ with respect to $A_i$ reads:
\begin{equation}
\label{eq:chi2_der}
\frac{\partial \chi^2 [A]}{\partial A_i} =
    2 \sum^{N}_{n = 1} \frac{K_{ni}\Delta_i}{\sigma^2_n}
    \left(\sum_m K_{nm} A_m \Delta_m - G_n\right).
\end{equation}

\section{Shannon-Jaynes entropy\label{sec:sj}}

In this appendix, the technical details for $S_{\text{SJ}}$, $S^{\pm}_{\text{SJ}}$, and the MaxEnt-SJ method are reviewed. Most of the equations presented in this appendix have been derived in Refs.~\cite{JARRELL1996133}, \cite{PhysRevB.96.155128}, and \cite{KAUFMANN2023108519}. We repeat the mathematical derivation here so that this paper is self-contained.   

\subsection{Entropy}

The Shannon-Jaynes entropy is defined as follows:
\begin{equation}
\label{eq:sj_std}
S_{\text{SJ}}[A] = \int d\omega
\left[
    A(\omega) - D(\omega) - A(\omega) \log \left(\frac{A(\omega)}{D(\omega)}\right)
\right],
\end{equation}
The discretization form of Eq.~(\ref{eq:sj_std}) is:
\begin{equation}
\label{eq:sj_dis}
S_{\text{SJ}}[A] = \sum_i \Delta_i
\left[
    A_i - D_i - A_i \log \left(\frac{A_i}{D_i} \right)
\right].
\end{equation}
The first derivative of $S_{\text{SJ}}[A]$ with respect to $A_i$ reads:
\begin{equation}
\label{eq:sj_der}
\frac{\partial S_{\text{SJ}}[A]}{\partial A_i} =
    - \Delta_i \log \left(\frac{A_i}{D_i}\right).
\end{equation}

\subsection{Parameterization of spectral function}

A singular value decomposition can be readily performed for the kernel function $K$:
\begin{equation}
K = U \xi V^{T},
\end{equation}
where $U$ and $V$ are column-orthogonal matrices and $\xi$ is the vector of singular values. So, the matrix element of $K$ reads:
\begin{equation}
\label{eq:Ksvd}
K_{ni} = \sum_m U_{nm} \xi_m V_{im}.
\end{equation}
The columns of $V$ can be understood as basis functions for the spectral function. That is to say they span a so-called singular value space. And in the singular value space the spectrum can be parameterized through:
\begin{equation}
\label{eq:Avu}
A_l = D_l \exp \left(\sum_m V_{lm} u_m\right).
\end{equation}
It is easy to prove that:
\begin{equation}
\label{eq:Avu_d}
\frac{\partial A_l}{\partial u_i} =
    D_l \exp\left(\sum_m V_{lm} u_m\right) V_{li} = A_l V_{li}.
\end{equation}

\subsection{Newton's method\label{sec:newton}}

Next, we will derive the equations for $\{u_m\}$. Let us recall the stationary condition [see also Eq.~(\ref{eq:opt})]:
\begin{equation}
\frac{\partial Q[A]}{\partial A_i} = 0.
\end{equation}
It is actually:
\begin{equation}
\label{eq:stationary}
\alpha \frac{\partial S_{\text{SJ}}[A]}{\partial A_i}
    -\frac{1}{2} \frac{\partial \chi^2[A]}{\partial A_i} = 0.
\end{equation}
Substituting Eqs.~(\ref{eq:sj_der}) and (\ref{eq:chi2_der}) into the above equation:
\begin{equation}
\alpha \Delta_i \log \left(\frac{A_i}{D_i}\right) +
    \sum^{N}_{n=1} \frac{K_{ni} \Delta_i}{\sigma^2_{n}}
    \left(\sum_m K_{nm} A_m \Delta_m - G_n \right) = 0.
\end{equation}
Eliminating $\Delta_i$:
\begin{equation}
\label{eq:stat_cond}
\alpha \log \left(\frac{A_i}{D_i}\right) +
    \sum^{N}_{n=1} \frac{K_{ni}}{\sigma^2_{n}}
    \left(\sum_m K_{nm} A_m \Delta_m - G_n \right) = 0.
\end{equation}
Substituting Eqs.~(\ref{eq:Avu}) and (\ref{eq:Ksvd}) into Eq.~(\ref{eq:stat_cond}):
\begin{eqnarray}
&~& \alpha \sum_m V_{im} u_m + \nonumber\\
&~&~~\sum^{N}_{n = 1} \frac{1}{\sigma^2_n} \sum_m U_{nm} \xi_m V_{im}
    \left(\sum_l K_{nl}A_l \Delta_l -G_n \right) = 0.
\end{eqnarray}
Now $\sum_m V_{im}$ can be removed from the two terms in the left-hand side. Finally we get:
\begin{equation}
\label{eq:um}
\alpha u_m + \sum^{N}_{n = 1} \frac{1}{\sigma^2_n}
    \xi_m U_{nm} \left(\sum_l K_{nl}A_l \Delta_l -G_n \right) = 0.
\end{equation}
Since $A_l$ depends on $\{u_m\}$ as well, Eq.~(\ref{eq:um}) is actually a non-linear equation about $\{u_m\}$. In the \texttt{ACFlow} package~\cite{Huang:2022}, the Newton's method is adopted to solve it. So we have to evaluate the following two variables and pass them to the Newton's algorithm~\cite{optimization_book}:
\begin{equation}
\label{eq:fm}
f_m = \alpha u_m +
    \xi_m \sum^{N}_{n = 1} \frac{1}{\sigma^2_n}
    U_{nm} \left(\sum_l K_{nl}A_l\Delta_l -G_n \right),
\end{equation}
\begin{equation}
\label{eq:Jmi}
J_{mi} = \frac{\partial f_m}{\partial u_i} =
    \alpha \delta_{mi} +
    \xi_m \sum^{N}_{n = 1} \frac{1}{\sigma^2_n}
    U_{nm} \sum_l K_{nl}A_l \Delta_l V_{li},
\end{equation}
where $J_{mi}$ can be considered as a Jacobian matrix. Note that Eq.~(\ref{eq:Avu_d}) is applied to derive Eq.~(\ref{eq:Jmi}).

Since the calculations for $f_m$ and $J_{mi}$ are quite complicated and time-consuming, the Einstein summation technology is used to improve the computational efficiency. The following three variables should be precomputed and stored during the initialization stage:
\begin{equation}
B_m = \sum^{N}_{n = 1} \frac{1}{\sigma^2_n} \xi_m U_{nm} G_n,
\end{equation}
\begin{equation}
W_{ml} = \sum_{pn} \frac{1}{\sigma^2_n}
    U_{nm}\xi_m U_{np} \xi_p V_{lp} \Delta_l D_l,
\end{equation}
\begin{equation}
W_{mli} = W_{ml} V_{li}.
\end{equation}
Clearly, the three variables only depend on the input Green's function $G$, the default model $D$, and the singular value decomposition of the kernel function $K$. Now $f_m$ and $J_{mi}$ can be reformulated in terms of them:
\begin{equation}
f_m = \alpha u_m + \sum_l W_{ml} w_l - B_m,
\end{equation}
\begin{equation}
J_{mi} = \alpha \delta_{mi} + \sum_l W_{mli} w_l,
\end{equation}
where
\begin{equation}
w_l = \exp \left(\sum_m V_{lm} u_m\right).
\end{equation}

\subsection{Positive-negative entropy}

For matrix-valued Green's function, the spectral functions of the off-diagonal components could exhibit negative weight. However, the ordinary MaxEnt method is only rigorous for non-negative spectrum~\cite{JARRELL1996133}. In order to remedy this problem, one could imagine that the off-diagonal spectral functions originate from a subtraction of two artificial positive functions~\cite{PhysRevB.96.155128},
\begin{equation}
A = A^{+} - A^{-},
\end{equation}
\begin{equation}
A^{-} = A^{+} - A.
\end{equation}
Assuming the independence of $A^+$ and $A^{-}$, the resulting entropy $S_{\text{SJ}}[A^+,A^-]$ is the sum of the respective entropies:
\begin{eqnarray}
\label{eq:s+-}
S_{\text{SJ}}[A^{+},A^{-}] =
&~&\int d\omega \left[A^{+} - D - A^{+} \log \left(\frac{A^{+}}{D}\right)\right] + \nonumber\\
&~&\int d\omega \left[A^{-} - D - A^{-} \log \left(\frac{A^{-}}{D}\right)\right].
\end{eqnarray}
Thus, $S_{\text{SJ}}[A^{+},A^{-}]$ is called the positive-negative entropy in the literature~\cite{PhysRevB.96.155128,KAUFMANN2023108519}. We at first eliminate $A^{-}$
\begin{widetext}
\begin{equation}
S_{\text{SJ}}[A,A^{+}] =
    \int d\omega \left[ A^{+} - D - A^{+} \log \left(\frac{A^{+}}{D}\right) \right]+
    \int d\omega \left [(A^{+} - A) - D - (A^{+} - A) \log \left(\frac{A^{+}-A}{D}\right)\right],
\end{equation}
\end{widetext}
and write:
\begin{equation}
Q[A,A^{+}] = \alpha S_{\text{SJ}}[A,A^{+}] - \frac{1}{2} \chi^{2}[A].
\end{equation}
Since we are searching for a maximum of $Q$ with respect to $A$ and $A^{+}$, we also apply:
\begin{equation}
\frac{\partial Q[A,A^{+}]}{\partial A^{+}} =
    \frac{\partial S_{\text{SJ}}[A,A^{+}]}{\partial A^{+}} = 0,
\end{equation}
to eliminate $A^{+}$. The following equations show some intermediate steps:
\begin{equation}
%\frac{\partial S_{\text{SJ}}[A,A^{+}]}{\partial A^{+}} =
\int d\omega
\left[
    \log \left(\frac{A^{+}}{D}\right)
    +\log \left(\frac{A^{+}-A}{D}\right)
\right]
= 0,
\end{equation}
\begin{equation}
\log \left(\frac{A^{+}}{D}\right) + \log \left(\frac{A^{+}-A}{D}\right) = 0,
\end{equation}
\begin{equation}
\frac{A^{+}(A^{+} - A)}{D^2} = 1.
\end{equation}
Finally, we obtain:
\begin{equation}
\label{eq:a+}
A^{+} = \frac{\sqrt{A^2 + 4D^2} + A}{2},
\end{equation}
\begin{equation}
\label{eq:a-}
A^{-} = \frac{\sqrt{A^2 + 4D^2} - A}{2}.
\end{equation}
Substituting Eqs.~(\ref{eq:a+}) and (\ref{eq:a-}) into Eq.~(\ref{eq:s+-}) to eliminate $A^{+}$ and $A^{-}$, we obtain:
\begin{eqnarray}
\label{eq:sj_pm2}
S_{\text{SJ}}[A^{+}, A^{-}] &=& \int d\omega
\Bigg[ \sqrt{A^2 + 4D^2} - 2D \nonumber \\
       &-&A\log\left(\frac{\sqrt{A^2 + 4D^2} + A}{2D}\right) \Bigg].
\end{eqnarray}
Its discretization form becomes:
\begin{eqnarray}
S_{\text{SJ}}[A^{+}, A^{-}] &=& \sum_m
\Bigg[ \sqrt{A_m^2 + 4D_m^2} - 2D_m \nonumber \\
&-& A_m\log\left(\frac{\sqrt{A_m^2 + 4D_m^2} + A_m}{2D_m}\right)
\Bigg]\Delta_m.
\end{eqnarray}
Note that Eq.~(\ref{eq:sj_pm2}) is exactly the same with Eq.~(\ref{eq:sj_pm}). $S_{\text{SJ}}[A^+,A^-]$ is just $S^{\pm}_{\text{SJ}}[A]$. The first derivative of $S_{\text{SJ}}[A^+,A^-]$ with respect to $A_i$ is:
\begin{equation}
\frac{\partial S_{\text{SJ}}[A^+,A^-]}{\partial A_i} = 
    -\Delta_i \log \left(\frac{A^+_i}{D_i}\right).
\end{equation}

\subsection{Parameterization for positive-negative spectral functions}

According to Eq.~(\ref{eq:a+}) and Eq.~(\ref{eq:a-}), we find:
\begin{equation}
A^{+} A^{-} = D^{2}.
\end{equation}
Inspired by Eq.~(\ref{eq:Avu}), it is naturally to parameterize $A^{+}$ and $A^{-}$ in the singular value space as well:
\begin{equation}
A^+_i = D_i \exp \left(\sum_m V_{im} u_m\right),
\end{equation}
\begin{equation}
A^-_i = D_i \exp \left(-\sum_m V_{im} u_m\right).
\end{equation}
Hence,
\begin{equation}
A_i = D_i \exp \left(\sum_m V_{im} u_m\right) - D_i \exp \left(-\sum_m V_{im} u_m\right) 
\end{equation}
\begin{equation}
A_i = D_i \left(w_i - \frac{1}{w_i}\right).
\end{equation}
\begin{equation}
\frac{\partial A^+_l}{\partial u_i} = A^+_l V_{li}.
\end{equation}
\begin{equation}
\frac{\partial A^-_l}{\partial u_i} = -A^-_l V_{li}.
\end{equation}
\begin{equation}
\frac{\partial A}{\partial u_i} = (A^+_l + A^-_l) V_{li}
    = D_l \left (w_l + \frac{1}{w_l} \right) V_{li}.
\end{equation}
By using the Einstein summation notations as defined in Section~\ref{sec:newton}, we immediately obtain:
\begin{equation}
f_m = \alpha u_m + \sum_l W_{ml}\left(w_l - \frac{1}{w_l}\right) - B_m,
\end{equation}
\begin{equation}
J_{mi} = \alpha \delta_{mi} +
    \sum_{l} W_{mli} \left(w_l + \frac{1}{w_l}\right).
\end{equation}

\section{Bayesian reconstruction entropy\label{sec:br}}

In this appendix, the technical details for $S_{\text{BR}}$, $S^{\pm}_{\text{BR}}$, and the MaxEnt-BR method are discussed.

\subsection{Entropy}

The Bayesian reconstruction entropy reads:
\begin{equation}
S_{\text{BR}}[A] = \int d\omega
\left[
    1 - \frac{A(\omega)}{D(\omega)} + \log \left(\frac{A(\omega)}{D(\omega)}\right)
\right].
\end{equation}
Its discretization form is:
\begin{equation}
S_{\text{BR}}[A] = \sum_i \Delta_i
\left[
    1 - \frac{A_i}{D_i} + \log \left(\frac{A_i}{D_i}\right)
\right].
\end{equation}
Its first derivative with respect to $A_i$ is:
\begin{equation}
\label{eq:br_der}
\frac{\partial S_{\text{BR}}[A]}{\partial A_i} = -\Delta_i
    \left(\frac{1}{D_i} - \frac{1}{A_i} \right).
\end{equation}

\subsection{Parameterization of spectral function}

The original parameterization of $A$ [see Eq.~(\ref{eq:Avu})] can not be used here. A new parameterization scheme for $A$ is necessary. Assumed that
\begin{equation}
\label{eq:br_ai}
\frac{1}{D_i} - \frac{1}{A_i} = \sum_m V_{im} u_m,
\end{equation}
then we have:
\begin{equation}
A_l = \frac{D_l}{ 1 - D_l \sum_m V_{lm} u_m}.
\end{equation}
So, the first derivative of $A_l$ with respect to $u_i$ reads:
\begin{equation}
\label{eq:br_ai_d}
\frac{\partial A_l}{\partial u_i} = A_l A_l V_{li}.
\end{equation}

\subsection{Newton's method}

Next, we would like to derive $f_m$ and $J_{mi}$ for the Bayesian reconstruction entropy. Substituting Eq.~(\ref{eq:br_der}) into Eq.~(\ref{eq:stationary}):
\begin{equation}
\alpha \Delta_i \left(\frac{1}{D_i} - \frac{1}{A_i}\right) +
    \sum^{N}_{n=1} \frac{K_{ni} \Delta_i}{\sigma^2_{n}}
    \left(\sum_m K_{nm} A_m \Delta_m - G_n \right) = 0.
\end{equation}
Eliminating $\Delta_i$:
\begin{equation}
\label{eq:br_newton}
\alpha \left(\frac{1}{D_i} - \frac{1}{A_i}\right) +
    \sum^{N}_{n=1} \frac{K_{ni}}{\sigma^2_{n}}
    \left(\sum_m K_{nm} A_m \Delta_m - G_n \right) = 0.
\end{equation}
Substituting Eq.~(\ref{eq:br_ai}) into Eq.~(\ref{eq:br_newton}):
\begin{eqnarray}
&~& \alpha \sum_m V_{im} u_m + \nonumber \\
&~&~~ \sum^{N}_{n = 1} \frac{1}{\sigma^2_n} \sum_m U_{nm} \xi_m V_{im}
      \left(\sum_l K_{nl}A_l \Delta_l -G_n \right) = 0.
\end{eqnarray}
Eliminating $\sum_m V_{im}$ again:
\begin{equation}
\label{eq:um2}
\alpha u_m + \sum^{N}_{n = 1} \frac{1}{\sigma^2_n}
    \xi_m U_{nm} \left(\sum_l K_{nl}A_l \Delta_l -G_n \right) = 0.
\end{equation}
Clearly, Eq.~(\ref{eq:um2}) is the same with Eq.~(\ref{eq:um}). Thus, the equations for $f_m$ and $J_{mi}$ are as follows:
\begin{equation}
\label{eq:fm2}
f_m = \alpha u_m + \xi_m \sum^{N}_{n = 1}
    \frac{1}{\sigma^2_n}
    U_{nm} \left(\sum_l K_{nl}A_l\Delta_l -G_n \right),
\end{equation}
\begin{equation}
\label{eq:Jmi2}
J_{mi} = \frac{\partial f_m}{\partial u_i} =
    \alpha \delta_{mi} +
    \xi_m \sum^{N}_{n = 1} \frac{1}{\sigma^2_n}
    U_{nm} \sum_l K_{nl}A_l A_l \Delta_l V_{li}.
\end{equation}
Note that Eq.~(\ref{eq:fm2}) is the same with Eq.~(\ref{eq:fm}). However, Eq.~(\ref{eq:Jmi2}) differs from Eq.~(\ref{eq:Jmi}) by a factor $A_l$ in the second term of the right-hand side. To derive Eq.~(\ref{eq:Jmi2}), Eq.~(\ref{eq:br_ai_d}) is used. Now the Einstein summation notation is adopted to simplify the calculations of $f_m$ and $J_{mi}$ again:
\begin{equation}
f_m = \alpha u_m + \sum_l W_{ml} w_l - B_m,
\end{equation}
\begin{equation}
J_{mi} = \alpha \delta_{mi} + \sum_l W_{mli} D_l w_l w_l,
\end{equation}
where
\begin{equation}
w_l = \frac{1}{1 - D_l \sum_m V_{lm} u_m}.
\end{equation}
Here, the definitions of $W_{ml}$, $W_{mli}$, and $B_m$ could be found in Section~\ref{sec:newton}.

\subsection{Positive-negative entropy}

Next, we would like to generalize the Bayesian reconstruction entropy to realize the positive-negative entropy algorithm to support the analytic continuation of matrix-valued Green's function~\cite{PhysRevB.96.155128}. The positive-negative entropy for the Bayesian reconstruction entropy is defined as follows:
\begin{eqnarray}
\label{eq:br_first}
S_{\text{BR}}[A^+, A^-] &=& \int d\omega
    \left[ 1 - \frac{A^+}{D} + \log\left(\frac{A^+}{D}\right) \right] \nonumber \\
&+& \int d\omega
    \left[ 1 - \frac{A^-}{D} + \log\left(\frac{A^-}{D}\right) \right].
\end{eqnarray}
Since $A = A^+ - A^-$, Eq.~(\ref{eq:br_first}) can be transformed into:
\begin{eqnarray}
\label{eq:br_second}
S_{\text{BR}}[A, A^+] &=& \int d\omega
    \left[ 1 - \frac{A^+}{D} + \log\left(\frac{A^+}{D}\right) \right] \nonumber \\
&+& \int d\omega
    \left[ 1 - \frac{A^+-A}{D} + \log\left(\frac{A^+-A}{D}\right) \right].
\end{eqnarray}
Then we should eliminate $A^{+}$ in Eq.~(\ref{eq:br_second}). Because
\begin{equation}
\frac{\partial S_{\text{BR}}[A, A^+] }{\partial A^+} = 0 =
    \int d\omega
    \left(\frac{1}{A^+} + \frac{1}{A^+-A} -\frac{2}{D} \right),
\end{equation}
we immediately get:
\begin{equation}
A^+ = \frac{\sqrt{A^2 + D^2} + D + A}{2},
\end{equation}
and
\begin{equation}
A^- = \frac{\sqrt{A^2 + D^2} + D - A}{2}.
\end{equation}
From the definitions of $A^{+}$ and $A^{-}$, it is easily to prove:
\begin{equation}
\label{eq:br_cons}
2A^+ A^- = (A^+ + A^-) D,
\end{equation}
and
\begin{equation}
\label{eq:br_cons1}
A^+ + A^- = \sqrt{A^2 + D^2} + D.
\end{equation}
\begin{widetext}
\noindent We would like to express $S_{\text{BR}}[A^+,A^-]$ in terms of $A$:
\begin{equation}
\label{eq:br_pm_orig}
S_{\text{BR}}[A^+, A^-]
= \int d\omega
\left[
    2 - \frac{A^+}{D} - \frac{A^-}{D} + \log\left(\frac{A^+A^-}{D^2}\right)
\right].
\end{equation}
Substituting Eq.~(\ref{eq:br_cons}) into Eq.~(\ref{eq:br_pm_orig}):
\begin{equation}
\label{eq:br_pm_orig1}
S_{\text{BR}}[A^+, A^-]
= \int d\omega
\left[
    2 - \frac{A^+ + A^-}{D} + \log\left(\frac{A^+ + A^-}{2D}\right)
\right].
\end{equation}
Substituting Eq.~(\ref{eq:br_cons1}) into Eq.~(\ref{eq:br_pm_orig1}):
\begin{equation}
\label{eq:br_pm2}
S_{\text{BR}}[A^+,A^-] = \int d\omega
\left[
    2 - \frac{\sqrt{A^2 + D^2} + D}{D} +
    \log \left(\frac{\sqrt{A^2 + D^2} + D}{2D}\right)
\right].
\end{equation}
Its discretization form reads:
\begin{equation}
S_{\text{BR}}[A^+,A^-] = \sum_m
\left[
    2 - \frac{\sqrt{A_m^2 + D_m^2} + D_m}{D_m} +
    \log \left(\frac{\sqrt{A_m^2 + D_m^2} + D_m}{2D_m}\right)
\right]\Delta_m.
\end{equation}
\end{widetext}
Note that Eq.~(\ref{eq:br_pm2}) is exactly the same with Eq.~(\ref{eq:br_pm}). $S_{\text{BR}}[A^+,A^-]$ is just $S^{\pm}_{\text{BR}}[A]$. Then the first derivative of $S_{\text{BR}}[A^+,A^-]$ with respect to $A_i$ is:
\begin{equation}
\frac{\partial S[A^+,A^-]}{\partial A_i} =
    -\Delta_i \left(\frac{1}{D_i} - \frac{1}{A_i^+}\right)
\end{equation}
This equation is very similar to Eq.~(\ref{eq:br_der}).

\subsection{Parameterization for positive-negative spectral functions}

Now we assume that:
\begin{equation}
A^+_l = \frac{D_l}{ 1 - D_l \sum_m V_{lm} u_m},
\end{equation}
and
\begin{equation}
A^-_l = \frac{D_l}{ 1 + D_l \sum_m V_{lm} u_m}.
\end{equation}
We also introduce:
\begin{equation}
w^+_l = \frac{1}{ 1 - D_l \sum_m V_{lm} u_m},
\end{equation}
and
\begin{equation}
w^-_l = \frac{1}{ 1 + D_l \sum_m V_{lm} u_m}.
\end{equation}
So $A^+_l = D_l w^+_l$, $A^-_l = D_l w^-_l$, and $A_l = D_l (w^+_l - w^-_l)$. It is easy to verify that the definitions of $A^+_l$ and $A^-_l$ obey Eq.~(\ref{eq:br_cons}). The first derivatives of $A^+_l$ and $A^-_l$ with respect to $u_i$ are:
\begin{equation}
\frac{\partial A^+_l}{\partial u_i} = A^+_l A^+_l V_{li},
\end{equation}
and
\begin{equation}
\frac{\partial A^-_l}{\partial u_i} = -A^-_l A^-_l V_{li}.
\end{equation}
The first derivative of $A_l$ with respect to $u_i$ is:
\begin{equation}
\frac{\partial A_l}{\partial u_i} = A^+_l A^+_l V_{li} + A^-_l A^-_l V_{li}.
\end{equation}
After some simple algebra, we easily get:
\begin{equation}
f_m = \alpha u_m + \sum_l W_{ml} (w^+_l - w^-_l) - B_m,
\end{equation}
\begin{equation}
J_{mi} = \alpha \delta_{mi} + \sum_l W_{mli} D_l (w^+_l w^+_l + w^-_l w^-_l).
\end{equation}

\bibliography{br}

\end{document}